\documentclass[english,12pt]{article}

\usepackage[LGR,T1]{fontenc}
\usepackage[utf8]{inputenc}
\usepackage{geometry}
\usepackage{array}
\usepackage{alltt}
\usepackage{bm}
\usepackage{float}
\usepackage{booktabs}
\usepackage{textcomp}
\usepackage{multirow}
\usepackage{amsmath}
\usepackage{amsthm}
\usepackage{amssymb}
\usepackage{authblk}
\usepackage{sectsty}
\usepackage{hyperref}
\usepackage{graphicx}
\usepackage{blindtext}
\usepackage{caption}
\usepackage{makecell}
\usepackage{subcaption}
\usepackage{lmodern}
\usepackage{multicol}
\usepackage{tabularray}
\usepackage{subfiles}
\usepackage{xr}

\usepackage{tikz}
\usetikzlibrary{arrows,automata}
\tikzset{
	semi/.style={
		semicircle,
		draw,
		minimum size=2em
	}
}

\usetikzlibrary{shapes, snakes, graphs, shapes.geometric, positioning}

\sectionfont{\fontsize{14}{13}\selectfont}
\subsectionfont{\fontsize{12}{12}\selectfont}

\allowdisplaybreaks

\makeatletter

\ProvideTextCommand{\~}{LGR}[1]{\char126#1}

\theoremstyle{plain}

\usepackage{amsmath}

\makeatother

\usepackage{babel}
\providecommand{\theoremname}{Theorem}

\usepackage[
backend=biber,
style=nature,
uniquename=init,
giveninits
]{biblatex}

\begin{document}

\title{Pseudo-value Based Mean Cumulative Count Regression}

\author[1]{Zachary R. McCaw}
\author[2$\dag$]{Alex Ocampo}
\author[3]{Enrico Giudice}
\author[4]{FengQi Song}
\author[4]{Jessica Gronsbell}

\affil[1]{University of North Carolina at Chapel Hill, NC, USA}
\affil[2]{Hoffman La Roche Ltd, Basel, Switzerland}
\affil[3]{Novartis Pharma AG, Basel, Switzerland}
\affil[4]{Department of Statistical Sciences, University of Toronto, Toronto, ON, Canada}

\affil[$\dag$]{Email: alex.ocampo@roche.com}
\setcounter{Maxaffil}{0}
\renewcommand\Affilfont{\itshape\small}

\date{}
\maketitle
\vspace{-11mm}
\begin{abstract}
The mean cumulative function (MCF) summarizes how events accumulate over time for a recurrent or multi-component endpoint. The MCF, and its integral over a given time horizon, the area under the MCF (AUMCF), provide interpretable summaries of recurrent-event burden in the presence of right-censoring and terminal events. Existing approaches for these estimands have focused primarily on nonparametric treatment comparisons, covariate-adjusted augmentation, and linearized test statistics. Herein, we propose a pseudo-value-based regression approach for estimating covariate effects on the MCF and AUMCF at a fixed truncation time. The proposed method uses influence-function-based pseudo-values as regression outcomes, allowing estimation with standard generalized estimating equation machinery and, under an identity link, ordinary least squares. Through simulation studies, we evaluate estimation accuracy, confidence interval coverage, type I error control, and power across a range of recurrent-event settings. We demonstrate the utility of the proposed covariate adjustment procedure through an application to the ORATORIO clinical trial, evaluating the safety and efficacy of ocrelizumab for the treatment of primary progressive multiple sclerosis. Overall, pseudo-value-based regression provides a simple and interpretable framework for modeling covariate effects on cumulative recurrent-event burden over time.
\end{abstract}


\section{Introduction}

Multiple and recurrent events arise frequently in biomedical applications. Examples include repeated hospitalizations, disease exacerbations, recurrent infections, and various cardiovascular events \cite{solomon2024finearts, anker2024reshape, bhatt2023dupilumab, feuerstadt2022ser109, lamas2024tact2}. In many settings, follow-up is partially censored and subject to the competing risk of death \cite{austin2016cr, mccaw2022cr}, which precludes observation of subsequent non-fatal events. This creates a setting in which the outcome of interest is not a single event-time, but rather the cumulative disease burden under censoring and truncation-by-death. Although an extensive literature exists for modeling recurrent event rates \cite{prentice1981regression, andersen1982cox, wei1989regression, lin2000semiparametric, ghosh2002marginal}, regression parameters from rate/intensity-based models are not always the quantities of primary scientific interest. In clinical applications, greater interest may lie in how treatment or patient characteristics affect the cumulative burden of recurrent events over a clinically relevant time horizon. \\

Two natural summaries of this burden are the mean cumulative function (MCF) \cite{ghosh2000nonparametric} and the area under the MCF (AUMCF) \cite{claggett2022nejm, gronsbell2026}. The MCF at time $\tau$ is the expected number of events of interest accumulated by time $\tau$ before the terminal event, while the AUMCF integrates this quantity over time and may be interpreted as a measure of total event burden, or equivalently the total event-free time lost, up to $\tau$. These functionals are appealing because they are directly interpretable on the original outcome scale and remain meaningful in the presence of a terminal event. In particular, they describe how frequently events occur and how long their cumulative consequences persist across follow-up. \\

Despite their interpretability, direct regression modeling of the MCF and AUMCF is less developed than regression modeling of recurrent event intensities, rates, or hazards. Parameters from intensity models describe the effect of treatment or covariates on the instantaneous event rate, but a direct translation into the effect on the cumulative number of events or integrated disease burden is typically lacking, especially when the recurrent event process is truncated by death. Moreover, the scientific question in many randomized trials is inherently marginal: how does treatment affect the expected cumulative disease burden by a fixed time point? This motivates regression procedures that target covariate effects on the MCF and AUMCF themselves rather than on the parameters of an underlying rate model. \\

Recent work on cumulative disease burden has focused more on nonparametric treatment comparisons and their corresponding covariate-adjusted estimators rather than on general regression modeling. For instance, Claggett et al.\ proposed a model-free, clinically interpretable summary of treatment effect for trials with multiple event-time outcomes \cite{claggett2022nejm}, while Gronsbell et al.\  developed nonparametric estimation and covariate-adjusted augmentation for the area under the mean cumulative function in the presence of terminal events \cite{gronsbell2026}. More recently, Sun et al.\  proposed a nonparametric covariate-adjusted estimator for the AUMCF in randomized trials to improve precision while preserving unconditional interpretability by linearizing the unadjusted test statistic \cite{sun2025}. These contributions underscore the appeal of cumulative burden estimands, but they do not directly provide a regression framework for modeling covariate effects on the MCF or AUMCF themselves. The present work complements this literature by targeting regression coefficients for these recurrent-event functionals at a fixed truncation time. \\

In this paper, we develop pseudo-value regression methods for the MCF and AUMCF at a given truncation time. We define the regression targets directly through population estimating equations, thereby separating the inferential targets from the parameters of any particular rate model for the recurrent events. Moreover, the use of pseudo-values allows the regression parameters to be estimated using standard generalized estimating equation (GEE) machinery. In the identity-link setting emphasized here, estimation reduces to ordinary least squares using pseudo-values as outcomes. We also show that the exact jackknife-based pseudo-values are well-approximated by their influence function-based counterparts, which reduces computational burden. With respect to empirical performance, we verify the estimation accuracy and confidence interval coverage of our proposal, as well as its type-I error and power, through extensive simulation studies. We also present a data application for disability accumulation as measured by three different time-to-event outcomes from the ORATORIO clinical trial of primary progressive multiple sclerosis. Together, these results support a simple and interpretable framework for regression analysis of cumulative recurrent event burden in the presence of death. Code implementing the proposed pseudo-value regression methods is available at \url{https://github.com/zrmacc/MCC}.\\



\section{Methods}

\subsection{Preliminaries}
\subsubsection{Setting \& Notation}

Consider the setting of a randomized clinical trial in which multiple events are potentially observed across time. Let $C$ denote the time of censoring and $D$ the time of a terminal event (e.g., death). Let $\bm{W}$ denote a baseline covariate vector, which includes the intercept and the treatment assignment. Define the latent recurrent-event counting process $N_{0}(t) = \sum_{k=1}^{\infty}\mathbb{I}(T_{k} \leq t)$, where $T_{k}$ is the time of the $k$th event and $\mathbb{I}(\cdot)$ is the indicator function. Let $N^{*}(t) = N_{0}(t \land D)$ denote the truncated-by-death process, which ceases to jump upon the occurrence of the terminal event. Note that the terminal event itself may or may not be counted as an event of interest. While the terminal event is not considered part of the recurrent-event process in the main development, including it simply requires adding a final jump to the event-counting process at time $D$. Denote by $N(t) = N^{*}(t \land C) = N_{0}(t \land D \land C)$ the observed counting process, which is subject to both right-censoring and truncation-by-death. We assume non-informative censoring in the sense that $C$ is independent of the latent recurrent-event process $N_{0}(\cdot)$, the terminal event time $D$, and the baseline covariates $\bm{W}$.

Each patient is observed from enrollment until the time of censoring or death, whichever occurs first. Let $X = C \land D$ denote the observation time. Define the status indicator as $\delta = \mathbb{I}(D \leq C)$ and the at-risk process as $Y(t) = \mathbb{I}(X \geq t)$. The observed data consist of $n$ independent and identically distributed tuples of the form $\big(N_{i}(\cdot), X_{i}, \delta_{i}, \bm{W}_{i}\big)$, where $i$ indexes the subject.


\subsubsection{Recurrent event functionals}

Define the mean cumulative function (MCF) as the expected value of the truncated-by-death process $\mu(t) = \mathbb{E}\{N^{*}(t)\}$. For a given truncation time $\tau$, $\mu(\tau)$ is the expected number of events occurring before death by time $\tau$. Define the area under the MCF (AUMCF) as:
\begin{align*}
    \alpha(\tau) = \mathbb{E}\left\{\int_{0}^{\tau}N^{*}(t)dt\right\} = \int_{0}^{\tau}\mu(t)dt
\end{align*}
The AUMCF can be interpreted as the total event-free time lost due to all events, and provides a description of the cumulative disease burden across the study period. \\

The MCF is expressible as $\mu(t) = \int_{0}^{t}S_{D}(u)dR(u)$ where $S_{D}(t) = \mathbb{P}(D \geq t)$ is the survival function for the time-to-death and $dR(t) = \mathbb{E}\{dN^{*}(t) | D \geq t\}$ is the expected increment conditional on survival to time $t$ \cite{ghosh2000nonparametric}. A consistent estimator for the MCF is:
\begin{align*}
    \widehat{\mu}(\tau) = \int_{0}^{\tau}\widehat{S}_{D}(u)\frac{d\overline{N}(u)}{\overline{Y}(u)},
\end{align*}
where $\widehat{S}_{D}$ is the Kaplan-Meier estimator of the survival function, $\overline{N}(t) = \sum_{i=1}^{n}N_{i}(t)$ is the sample-level counting process, and $\overline{Y}(t) = \sum_{i=1}^{n}Y_{i}(t)$ is the sample-level at-risk process. The plug-in estimator of the AUMCF is:
\begin{align*}
    \widehat{\alpha}(\tau) = \int_{0}^{\tau}\widehat{\mu}(t)dt = \int_{0}^{\tau}\int_{0}^{t}\widehat{S}_{D}(u)\frac{d\overline{N}(u)}{\overline{Y}(u)} = \int_{0}^{\tau}(\tau - u)\widehat{S}_{D}(u)\frac{d\overline{N}(u)}{\overline{Y}(u)}.
\end{align*}
For exposition, we let $\theta(\tau)$ denote a generic marginal estimand, where $\theta(\tau)$ may represent either the MCF $\mu(\tau)$ or the AUMCF $\alpha(\tau)$, and $\widehat \theta(\tau)$ its corresponding plug-in estimator.

\subsection{Pseudo-value regression}

\subsubsection{Regression targets}

We next consider regression modeling for the MCF and AUMCF. For a fixed truncation time $\tau$, define the covariate-specific estimands:
\begin{align*}
    \mu(\tau \mid \bm{W}_{i}) &= \mathbb{E}\{N_{i}^{*}(\tau) \mid \bm{W}_{i}\}, \\
    \alpha(\tau \mid \bm{W}_{i}) &= \mathbb{E}\left\{\int_{0}^{\tau} N_{i}^{*}(u)du \,\middle|\, \bm{W}_{i}\right\}.
\end{align*}
These correspond to the conditional mean number of events by time $\tau$ and the conditional cumulative event-time burden up to time $\tau$, respectively. We similarly let $\theta(\tau \mid \bm{W}_i)$ denote a generic conditional estimand throughout our subsequent discussion.

The conditional quantity $\theta(\tau \mid \bm{W}_i)$ need not be exactly linear in $\bm{W}$. We therefore define the regression coefficients as working-model parameters through population estimating equations, rather than as parameters of an underlying recurrent-event intensity model. For the identity-link models emphasized here, we define $\bm{\beta}_0$ as the unique solution to:
\begin{align*}
    \mathbb{E}\left[\bm{W}\left\{\theta(\tau \mid \bm{W}) - \bm{W}^{\top}\bm{\beta}_{0}\right\}\right] &= \bm{0}.
\end{align*}
When the linear mean model is correctly specified, these coefficients have the usual conditional-mean interpretation. Otherwise, they represent the best linear projection of the conditional MCF or AUMCF onto the chosen covariates. To estimate these regression targets, we use pseudo-values as subject-specific outcomes.

\subsubsection{Pseudo-values}
Pseudo-values are synthetic subject-specific outcomes whose sample mean targets the same effect as the full-sample plug-in estimator of a population-level functional \cite{andersen2003, andersen2010}. Lemma 1 in Supplementary Section 2 provides a proof of this result. The exact jackknife pseudo-value for subject $i$ is:
\begin{align}
    \widehat{\xi}_{i}(\tau) = n\widehat{\theta}(\tau) - (n-1)\widehat{\theta}^{(-i)}(\tau),
    \label{eqn:exact-pseudo-value}
\end{align}
where $\widehat{\theta}^{(-i)}(\tau)$ is the estimator computed after removing subject $i$. Because censoring and truncation by death are incorporated into their construction, pseudo-values enable subsequent regression modeling using the standard machinery of GEEs.

\subsubsection{Influence-function pseudo-values}

Direct calculation of the jackknife pseudo-values requires recomputing the estimator after removal of each subject, which can become computationally burdensome. This is especially relevant when the estimand must be evaluated repeatedly across simulation settings or candidate truncation times. We therefore approximate the pseudo-values using the infinitesimal jackknife, replacing leave-one-out recalculation with an empirical influence function \cite{efron1982jackknife}. This approach was recently developed to perform interpretable mediation analysis for key survival estimands, including the survival probability, restricted mean survival time, and cumulative incidence \cite{ocampo2024}.

Suppose that the estimator $\widehat{\theta}(\tau)$ admits an influence function expansion:
\begin{align*}
    \widehat{\theta}(\tau) - \theta(\tau) = \frac{1}{n}\sum_{i=1}^{n}\varphi_{i}(\tau) + o_{p}(n^{-1/2}),
\end{align*}
where $\varphi_{i}(\tau)$ is the influence function contribution for subject $i$. Let $\widehat{\varphi}_{i}(\tau)$ denote the corresponding empirical influence function obtained by replacing unknown parameters with their consistent estimators. We show in Lemma 1 in Supplementary Section 2 that the jackknife pseudo-values are asymptotically equivalent to: 
\begin{align}
    \widetilde{\xi}_{i}(\tau) = \widehat{\theta}(\tau) + \widehat{\varphi}_{i}(\tau).
    \label{eqn:approx-pseudo-value}
\end{align}
Accordingly, we use $\widetilde{\xi}_{i}(\tau)$ as an approximation to $\widehat{\xi}_{i}(\tau)$ in all regression analyses. This approximation avoids recomputation of the estimator $n$ times for each truncation time $\tau$.

Influence function expansions for the plug-in estimators of the MCF and AUMCF have previously been derived \cite{ghosh2000nonparametric, gronsbell2026}; we state them here for completeness. The influence function for the MCF is:
\begin{align*}
    \varphi_{i}(t) = \int_{0}^{t}\frac{\mu(u)}{y(u)}dM_{D,i}(u) - \mu(t)\int_{0}^{t}\frac{1}{y(u)}dM_{D,i}(u) + \int_{0}^{t}\frac{S_{D}(u)}{y(u)}dM_{i}(u),
\end{align*}
where $\mu(t) = \mathbb{E}\{N^{*}(t)\}$, $y(t) = \mathbb{P}(X \geq t)$, $M_{i}(t) = N_{i}(t) - \int_{0}^{t}Y_{i}(u)dR(u)$ is the recurrent-events residual process, and $M_{D,i}(t) = N_{D,i}(t) - \int_{0}^{t}Y_{i}(u)d\Lambda_{D}(u)$ is the terminal-event martingale. Regarding the latter, $N_{D,i}(t) = \mathbb{I}(X_{i} \leq t, \delta_{i} = 1)$ is the counting process for the terminal event only, and $\Lambda_{D}(t)$ is the cumulative hazard for this process. The influence function for the AUMCF is:
\begin{align*}
    \varphi_{i}(\tau) = \int_{0}^{\tau}\frac{(\tau-u)S_{D}(u)}{y(u)}dM_{i}(u) - \int_{0}^{\tau}\frac{\int_{u}^{\tau}(\tau-s)d\mu(s)}{y(u)}dM_{D,i}(u).
\end{align*}
The empirical influence functions $\widehat{\varphi}_{i}(\tau)$ are obtained by substituting consistent estimators for the unknown population quantities in these expressions.

\subsubsection{Estimation via generalized estimating equations}

For both the MCF and AUMCF, we consider the working model
\begin{align*}
    \theta\left(\tau \mid \bm{W} \right) = g^{-1}(\bm{W}^{\top}\bm{\beta}).
\end{align*}
The estimator of the regression parameter $\widehat{\bm{\beta}}$ is the solution to the following estimating equation with pseudo-values as outcomes:
\begin{align*}
    \mathcal{U}_n(\bm{\beta}) = n^{-1}\sum_{i=1}^{n} \bm{D}_{i}(\bm{\beta}) v_{i}^{-1} \left\{ \widetilde{\xi}_{i}(\tau) - g^{-1}( \bm{W}_{i}^{\top}\bm{\beta}) \right\} = \bm{0},
\end{align*}
where $\bm{D}_{i}(\bm{\beta}) = \left\{\frac{\partial g^{-1}( \bm{W}_{i}^{\top}\bm{\beta})}{\partial \bm{\beta}}\right\}$ and $v_{i}$ is a working variance. For estimating the standard errors of $\widehat{\bm{\beta}}$, the robust variance estimator is $\widehat{\mathbb{V}}(\widehat{\bm{\beta}}) = \bm{A}^{-1}\bm{B}\bm{A}^{-1}$ where:
\begin{align*}
    \bm{A}(\bm{\beta}) 
    &= 
    \sum_{i=1}^{n}\bm{D}_{i}(\bm{\beta})v_{i}^{-1}\bm{D}_{i}^{\top}(\bm{\beta}), \\
    \bm{B}(\bm{\beta}) 
    &= 
    \sum_{i=1}^{n}\bm{D}_{i}(\bm{\beta})
    \frac{\left\{ \widetilde{\xi}_{i}(\tau) - g^{-1}( \bm{W}_{i}^{\top}\bm{\beta}) \right\}^{2}}{v_{i}^{2}}
    \bm{D}_{i}^{\top}(\bm{\beta}). 
\end{align*}
Unless otherwise specified, we adopt the identity link $g(\eta) = \eta$ and subject-agnostic working variance $v_{i} \equiv 1$. In this case, the estimating equation model reduces to standard linear regression and can be estimated via ordinary least squares. The usual model-based variance estimator is:
\begin{align*}
    \widehat{\mathbb{V}}_{\mathrm{MB}}(\widehat{\bm{\beta}}) 
    = \widehat{\sigma}^{2}\left(\sum_{i=1}^{n}\bm{W}_{i}\bm{W}_{i}^{\top}\right)^{-1},
\end{align*}
where $\widehat{\sigma}^{2}$ is the estimated residual variance from the pseudo-value regression.

\section{Simulation Studies}


\subsection{Simulation methods}

We consider the setting of a two-armed randomized clinical trial with a binary treatment indicator $A \in \{0, 1\}$ and a continuous baseline covariate $Z \sim N(0, 1)$, independent of treatment assignment. The primary estimands are the treatment and covariate effects on functionals of the recurrent-events process, namely the MCF and AUMCF. 

\subsubsection{Data generation}

Data for two treatment arms of equal size $N$ were simulated for $N \in \{50, 100, 200\}$. Recurrent events were simulated from subject-specific, time-homogeneous Poisson processes with truncation-by-death and both random and administrative censoring. For each subject $i$, let $A_{i}$ denote the treatment assignment and $Z_{i}$ the covariate. Random censoring $\tilde{C}_{i}$ and death $D_{i}$ times were simulated from independent exponential distributions with respective rate parameters $\lambda_{C}$ and $\lambda_{D}$. Unless otherwise specified, $\lambda_{C} = 0.25$ and $\lambda_{D} = 0.25$. The study duration was set to $C_{0} = 4$ years, and truncation times of $\tau \in \{2, 3, 4\}$ were considered. Let $C_{i} = \min(\tilde{C}_{i}, C_{0})$ denote the effective censoring time, and $X_{i} = \min(C_{i}, D_{i})$ the observation time. Let $T_{i,l}$ denote the latent arrival time for subject $i$'s $l$th event, and let $\Delta_{i,l}$ denote the $l$th latent gap time, with $T_{i,l} = \sum_{l'=1}^{l}\Delta_{i,l'}$. Gap times were simulated independently from exponential distributions with subject-specific event rates:
\begin{align*}
\lambda_{E,i} = \lambda_{0}\exp\left(\alpha_{A}A_{i} + \alpha_{Z}Z_{i} + \alpha_{AZ}A_{i}Z_{i}\right).
\end{align*}
Unless otherwise specified, $\lambda_{0} = 1$. The rate parameters $(\alpha_{A}, \alpha_{Z}, \alpha_{AZ})$ in the generative model were set to modulate the treatment, covariate, and interaction effects on the recurrent-event rate; these are distinct from the regression coefficients, $\beta$ and $\gamma$, that are the targets of inference. Each rate parameter was varied across the set $\{\ln(0.75), \ln(1.00), \ln(1.25)\}$ while fixing the remaining parameters at zero. 

To assess performance when covariates affect the terminal-event process, we additionally considered simulations in which the recurrent-event rate was held fixed at $\lambda_{E,i} = \lambda_{0}$, while the terminal-event rate was varied according to:
\begin{align*}
\lambda_{D,i} = \lambda_{D}\exp\left(\delta_{A}A_{i} + \delta_{Z}Z_{i} + \delta_{AZ}A_{i}Z_{i}\right).
\end{align*}
The terminal-event parameters $(\delta_{A}, \delta_{Z}, \delta_{AZ})$ were varied one at a time while fixing the remaining terminal-event rate parameters at zero.

For frailty simulations, dependence was induced between the death and recurrent-event rates via a shared multiplicative frailty. Frailties $\lambda_{F,i}$ were generated independently from a gamma distribution with $\mathbb{E}(\lambda_{F,i}) = 1$ and variance $\mathbb{V}(\lambda_{F,i}) = \sigma_{F}^{2}$, where $\sigma_{F}^{2}$ is a tunable hyper-parameter that controls the strength of dependence. The final per-subject death and recurrent-event rates were then set to $\lambda_{D,i} = \lambda_{D}\lambda_{F,i}$ and $\lambda_{E,i}^{*} = \lambda_{E,i}\lambda_{F,i}$. When $\sigma_{F}^{2} = 0$, $\lambda_{F,i} \equiv 1$ for all subjects, and the baseline data-generating process is recovered.

\subsubsection{Targets of inference}
For a given truncation time $\tau$ and design vector $\bm{W}_{i} = (1, A_{i}, Z_{i}, A_{i}Z_{i})^{\top}$, define $\bm{\beta}$ as the unique solution to the population estimating equation:
\begin{align*}
    \mathbb{E}\left[\bm{W}_{i}\left\{\mu(\tau \mid \bm{W}_{i}) - \bm{W}_{i}^{\top}\bm{\beta}\right\}\right] = \bm{0},
\end{align*}
where $\mu(\tau \mid \bm{W}_{i}) = \mathbb{E}\{N_{i}^{*}(\tau) \mid \bm{W}_{i}\}$. Similarly, for the AUMCF, define $\bm{\gamma}$ as the unique solution to:
\begin{align*}
    \mathbb{E}\left[\bm{W}_{i}\left\{\alpha(\tau \mid \bm{W}_{i}) - \bm{W}_{i}^{\top}\bm{\gamma}\right\}\right] = \bm{0},
\end{align*}
where $\alpha(\tau \mid \bm{W}_{i}) = \mathbb{E}\left\{\int_{0}^{\tau} N_{i}^{*}(u)du \,\middle|\, \bm{W}_{i}\right\}$. 

\subsubsection{Ground truth}
The target coefficients depend on the truncation time $\tau$ and on the parameters of the data-generating mechanism, including both the recurrent and terminal event rates. Ground truth values for the targets of inference were approximated by Monte Carlo simulation using $R = 100$ data sets with sample size $N = 10{,}000$ per arm in the absence of random censoring. Within each data set, the target coefficients were approximated by solving the sample analogue of the population estimating equations. The reported ground truth values were taken as the mean of these coefficient estimates across the $R$ replicates.


\subsection{Simulation results}


\subsubsection{Pseudo-value validation}

\begin{figure}[ht]
    \centering
    \begin{subfigure}{0.48\textwidth}
        \centering
        \caption{MCF}
        \includegraphics[width=\textwidth]{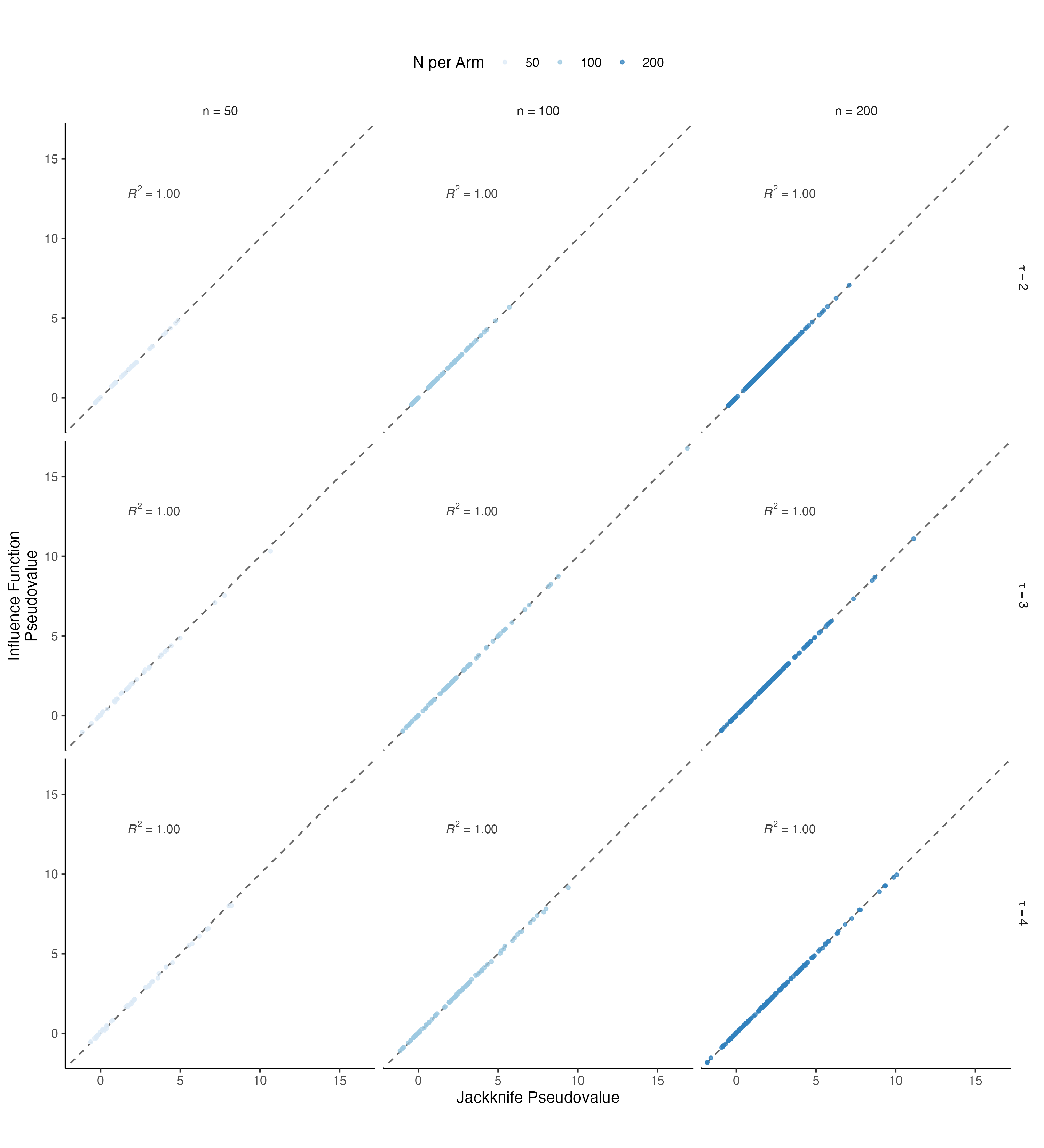}
    \end{subfigure}
    \begin{subfigure}{0.48\textwidth}
        \centering
        \caption{AUMCF}
        \includegraphics[width=\textwidth]{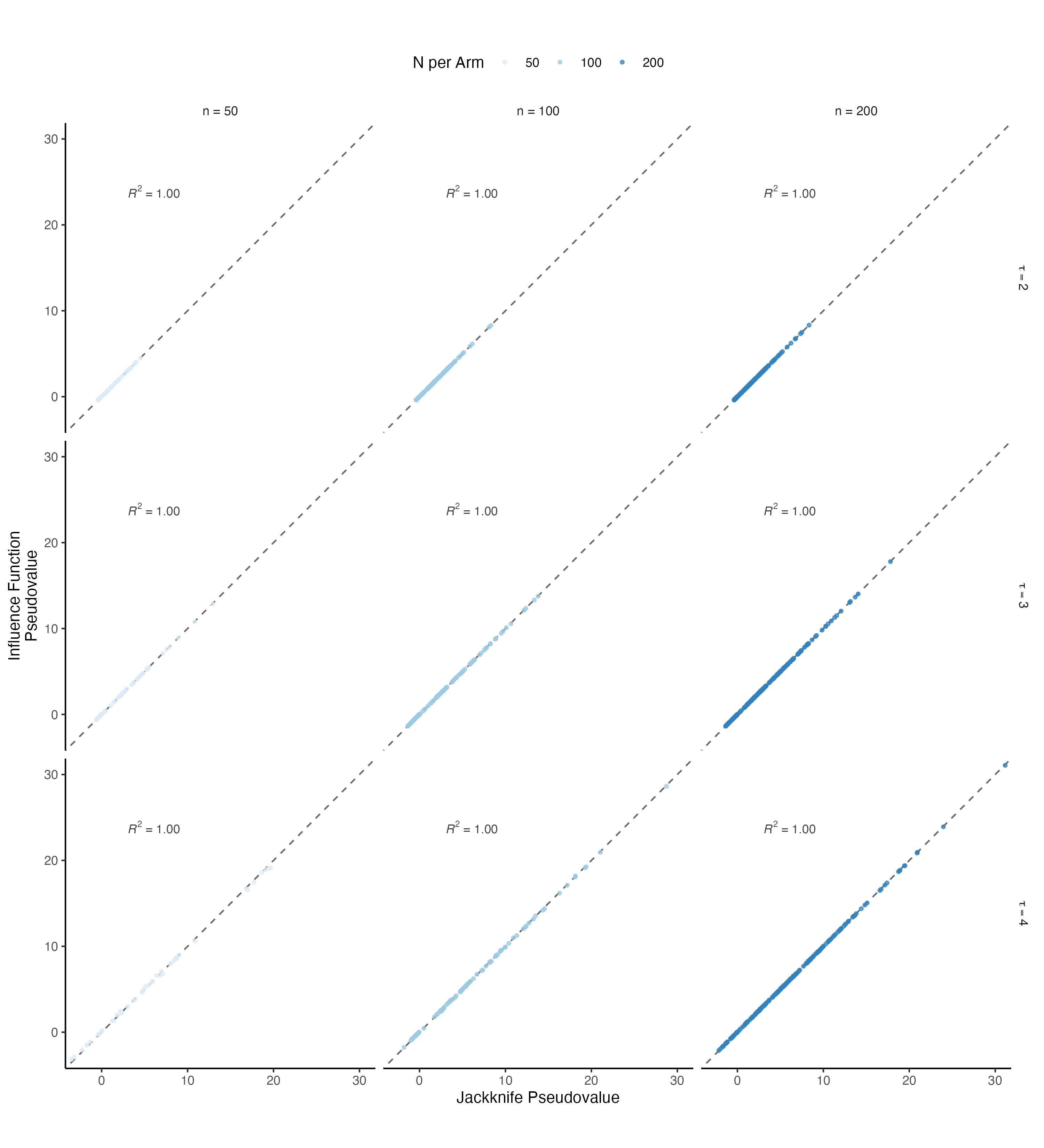}
    \end{subfigure}
    \caption{\textbf{Jackknife and influence function-based pseudo-values are concordant.} Across various sample sizes and truncation times, pseudo-values were calculated either exactly, via the leave-one-subject-out jackknife, or approximately, via the influence function.}
    \label{fig:concord}
\end{figure}

We first verified the influence function-based calculation of the pseudo-values. For simplicity, recurrent events data were simulated in the absence of any treatment or covariate effects. Pseudo-values were calculated in two ways, either exactly via the leave-one-subject-out jackknife, as described in equation (\ref{eqn:exact-pseudo-value}), or approximately via the influence function, as described in equation (\ref{eqn:approx-pseudo-value}). The scatter plots in \textbf{Figure \ref{fig:concord}} directly compare these two approaches. Although the methods are not numerically identical, the approximation is sufficiently accurate for practical use. Within all strata, the $R^{2}$ was 1.00 to two significant digits for both the MCF and AUMCF. The mean absolute difference was 0.01, or 0.4\% of the mean absolute jackknife pseudo-value of 2.4 (across all strata). 

\subsubsection{Baseline estimation accuracy and coverage}

To assess estimation accuracy and confidence interval coverage, data were generated by varying the effects of the treatment assignment $A$, the covariate $Z$, and their interaction $AZ$, on the recurrent event rate. Rate parameters were varied one at a time, fixing the remaining parameters to zero. Our focus throughout is on estimating the corresponding regression coefficients defined by the population estimating equations rather than on recovery of the rate-model parameters themselves. The target parameters were estimated via linear regression of the pseudo-values on $\bm{W} = (1, A, Z, AZ)^{\top}$. Confidence interval coverage was evaluated for both model-based and robust standard errors (SEs). For the latter, the leverage-adjusted heteroskedasticity-consistent HC3 standard errors recommended by Long and Ervin \cite{long2000} were employed. Results are shown from $2000$ simulation replicates per generative scenario. 

\textbf{Table \ref{tab:robust-ses-beta}} compares the model-based and robust SEs for estimating the effect of treatment on the MCF ($\beta_{A}$) with respect to the root-mean-square magnitude and coverage. Coverage is defined as the fraction of simulation replicates in which the 95\% confidence interval included the ground truth. As expected, the robust SEs were generally larger than the model-based SEs and provided slightly higher coverage. Given the similar performance of the two approaches in these simulations, we report model-based SEs hereafter for simplicity. Analogous results for the effect of treatment on the AUMCF ($\gamma_{A}$) are available in \textbf{Table \ref{supp:tab:robust-ses-gamma}}, and the behavior for other regression coefficients was qualitatively similar.  

\begin{figure}[ht]
    \centering
    \begin{subfigure}{0.48\textwidth}
        \centering
        \caption{MCF}
        \includegraphics[width=\textwidth]{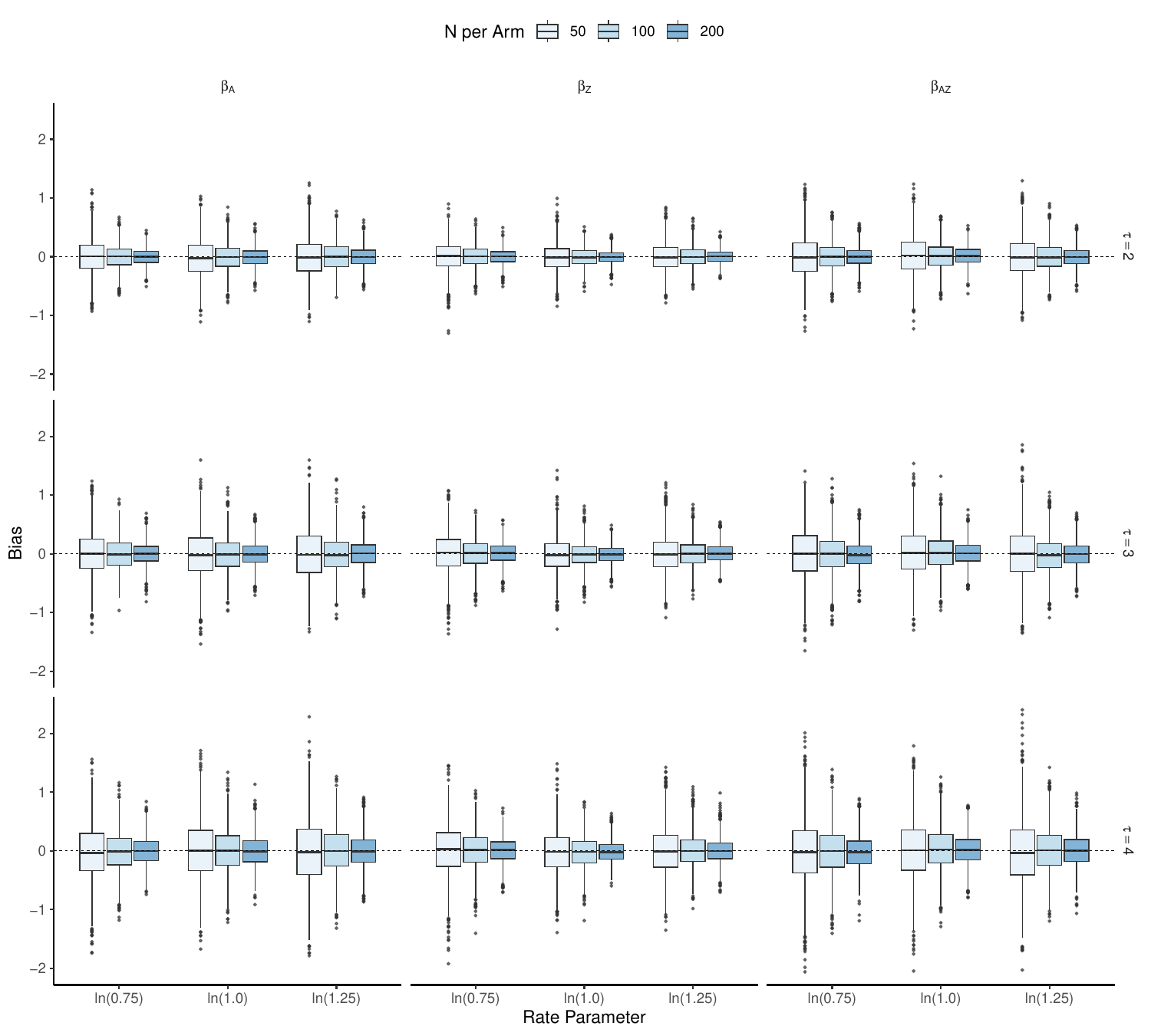}
    \end{subfigure}
    \begin{subfigure}{0.48\textwidth}
        \centering
        \caption{AUMCF}
        \includegraphics[width=\textwidth]{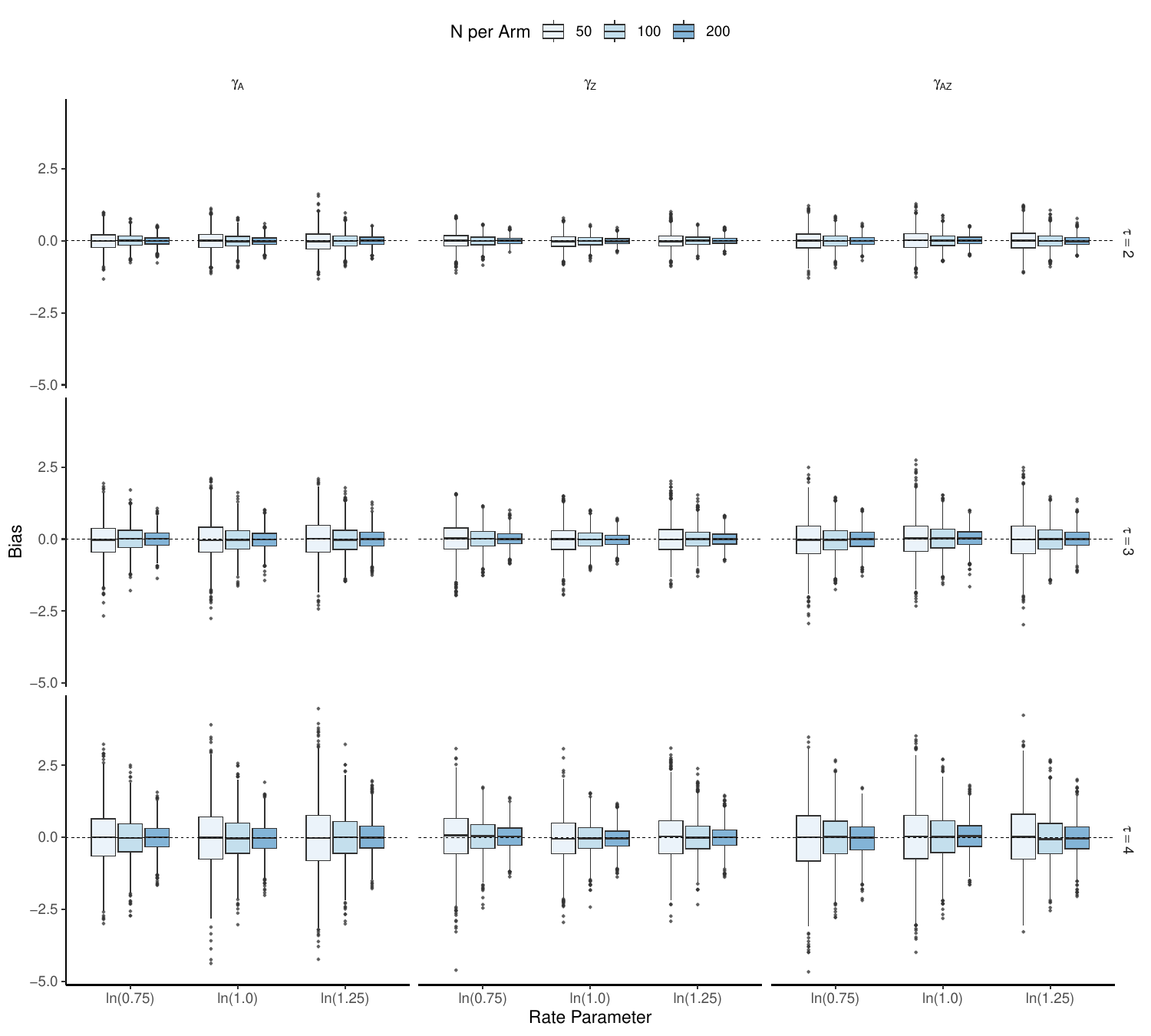}
    \end{subfigure}
    \caption{\textbf{Estimation of target regression parameters is unbiased.} Estimation error is shown for the MCF regression coefficients $\bm{\beta}$ and AUMCF regression coefficients $\bm{\gamma}$ across sample sizes, truncation times, and recurrent-event rate parameters. Rate parameters were varied one at a time while the remaining parameters were fixed at zero.}
    \label{fig:bias}
\end{figure}

\begin{figure}[ht]
    \centering
    \begin{subfigure}{0.48\textwidth}
        \centering
        \caption{MCF}
        \includegraphics[width=\textwidth]{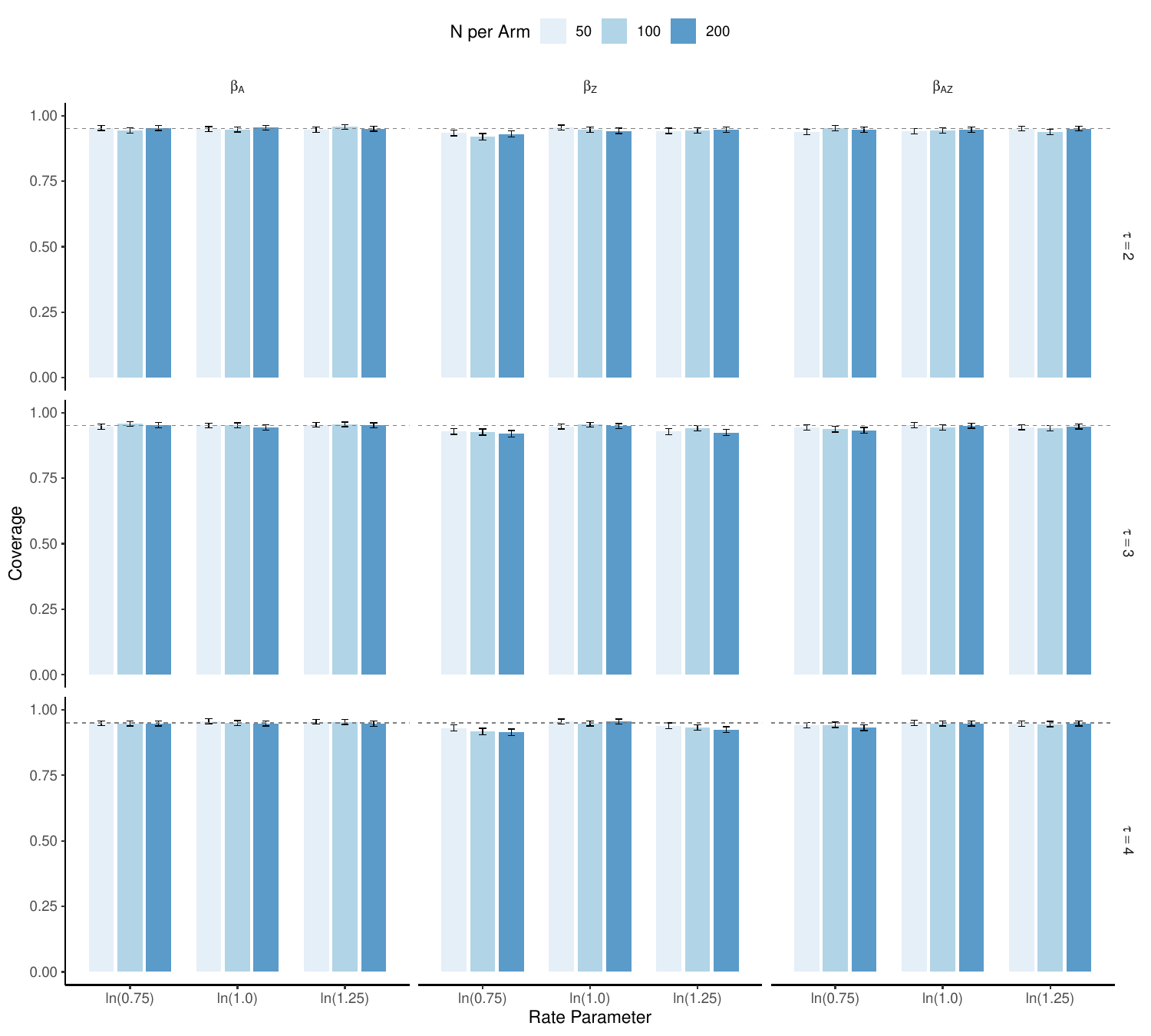}
    \end{subfigure}
    \begin{subfigure}{0.48\textwidth}
        \centering
        \caption{AUMCF}
        \includegraphics[width=\textwidth]{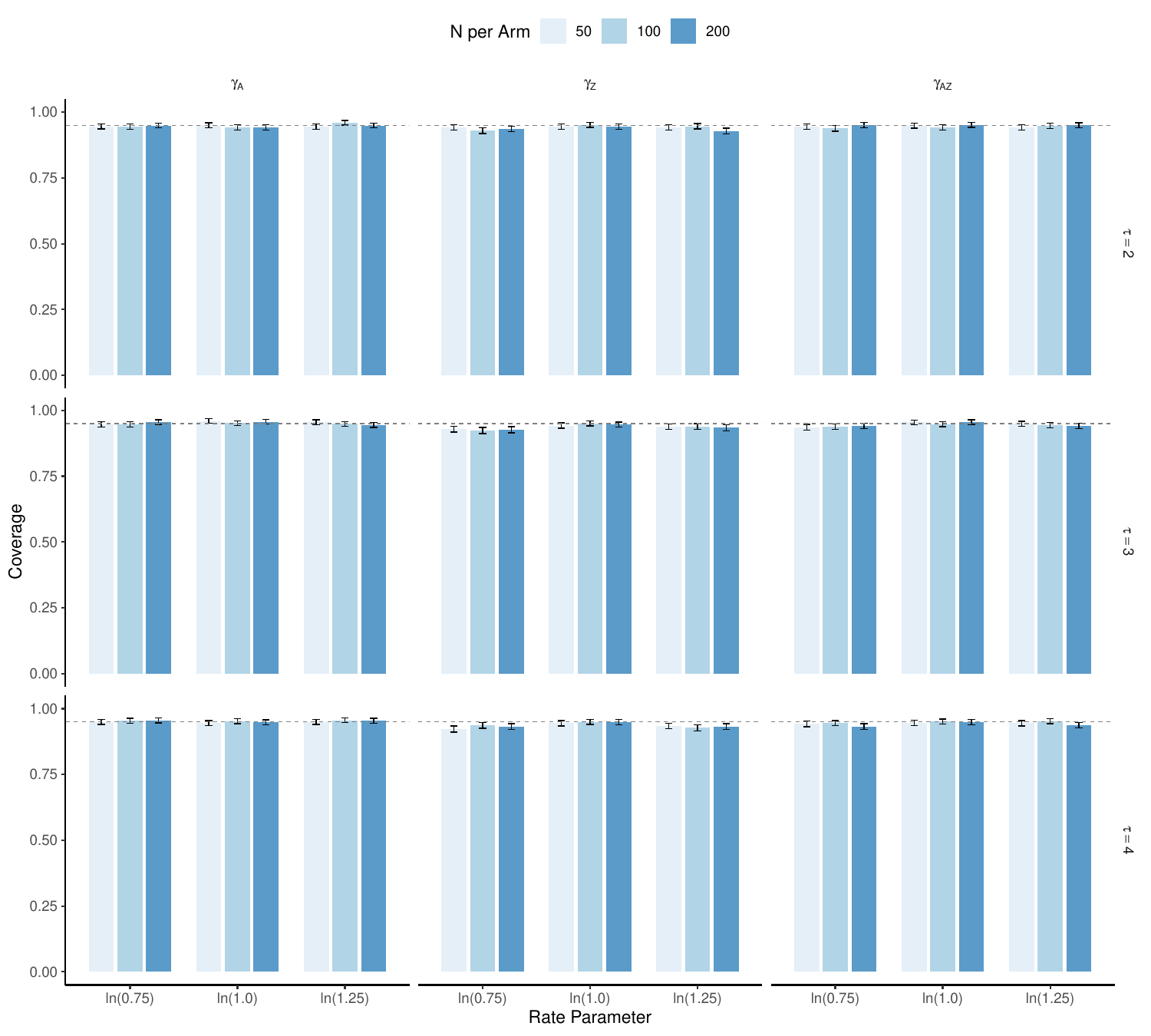}
    \end{subfigure}
    \caption{\textbf{Confidence intervals exhibit nominal coverage.} Empirical coverage of 95\% confidence intervals is shown for the MCF regression coefficients $\bm{\beta}$ and AUMCF regression coefficients $\bm{\gamma}$ across sample sizes, truncation times, and recurrent-event rate parameters. Confidence intervals were constructed using model-based standard errors.}
    \label{fig:coverage}
\end{figure}

\textbf{Figure \ref{fig:bias}} presents box plots of the estimation error, $\hat{\bm{\beta}} - \bm{\beta}$ for the MCF and $\hat{\bm{\gamma}} - \bm{\gamma}$ for the AUMCF. The estimation error is centered at zero, indicating the estimation of the population regression coefficients is unbiased. The decreasing width with increasing sample size illustrates the expected trend towards increasing precision. \textbf{Figure \ref{fig:coverage}} presents bar plots for the empirical coverage of 95\% confidence intervals for all population regression coefficients, using model-based SEs. The coverage is approximately nominal in all cases.  

\begin{table}
\centering 
\begin{tabular}{lrr|rr|rr}
\toprule
\makecell{Sample\\Size} & 
\makecell{Truncation\\Time} & 
\makecell{Rate\\Parameter} &
\makecell{Model-based\\SE} &
\makecell{Model-based\\Coverage} &
\makecell{Robust\\SE} &
\makecell{Robust\\Coverage} \\
\midrule
50 & 2 & ln(0.75) & 0.2926 & 0.944 & 0.2990 & 0.950 \\
50 & 2 & ln(1.0) & 0.3157 & 0.951 & 0.3226 & 0.956 \\
50 & 2 & ln(1.25) & 0.3406 & 0.954 & 0.3484 & 0.961 \\
50 & 3 & ln(0.75) & 0.3873 & 0.958 & 0.3954 & 0.960 \\
50 & 3 & ln(1.0) & 0.4196 & 0.956 & 0.4292 & 0.959 \\
50 & 3 & ln(1.25) & 0.4547 & 0.954 & 0.4650 & 0.957 \\
50 & 4 & ln(0.75) & 0.4781 & 0.946 & 0.4885 & 0.951 \\
50 & 4 & ln(1.0) & 0.5180 & 0.952 & 0.5292 & 0.958 \\
50 & 4 & ln(1.25) & 0.5668 & 0.949 & 0.5792 & 0.953 \\ \hline
100 & 2 & ln(0.75) & 0.2062 & 0.949 & 0.2084 & 0.953 \\
100 & 2 & ln(1.0) & 0.2230 & 0.947 & 0.2253 & 0.951 \\
100 & 2 & ln(1.25) & 0.2401 & 0.951 & 0.2426 & 0.954 \\
100 & 3 & ln(0.75) & 0.2716 & 0.955 & 0.2746 & 0.958 \\
100 & 3 & ln(1.0) & 0.2948 & 0.951 & 0.2979 & 0.952 \\
100 & 3 & ln(1.25) & 0.3198 & 0.947 & 0.3232 & 0.948 \\
100 & 4 & ln(0.75) & 0.3365 & 0.956 & 0.3400 & 0.959 \\
100 & 4 & ln(1.0) & 0.3664 & 0.951 & 0.3702 & 0.952 \\
100 & 4 & ln(1.25) & 0.4002 & 0.949 & 0.4044 & 0.950 \\ \hline
200 & 2 & ln(0.75) & 0.1460 & 0.953 & 0.1467 & 0.954 \\
200 & 2 & ln(1.0) & 0.1574 & 0.945 & 0.1582 & 0.945 \\
200 & 2 & ln(1.25) & 0.1692 & 0.949 & 0.1701 & 0.949 \\
200 & 3 & ln(0.75) & 0.1922 & 0.943 & 0.1931 & 0.943 \\
200 & 3 & ln(1.0) & 0.2079 & 0.949 & 0.2089 & 0.950 \\
200 & 3 & ln(1.25) & 0.2258 & 0.939 & 0.2270 & 0.941 \\
200 & 4 & ln(0.75) & 0.2381 & 0.954 & 0.2393 & 0.955 \\
200 & 4 & ln(1.0) & 0.2590 & 0.943 & 0.2603 & 0.944 \\
200 & 4 & ln(1.25) & 0.2830 & 0.952 & 0.2845 & 0.953 \\
\bottomrule
\end{tabular}
\caption{\textbf{Model-based standard errors provide adequate coverage in the evaluated settings.} The rate parameter is $\alpha_{A}$, the effect of treatment on the recurrent event rate. The target of inference is $\beta_{A}$, the effect of treatment on the mean cumulative function. The model-based and robust standard errors are the root-mean-square across simulation replicates. Coverage is the fraction of simulations in which the 95\% confidence interval included the ground truth.}
\label{tab:robust-ses-beta}
\end{table}


\subsubsection{Frailty study}
\begin{figure}[ht]
    \centering
    \begin{subfigure}{0.48\textwidth}
        \centering
        \caption{Estimation accuracy}
        \includegraphics[width=\textwidth]{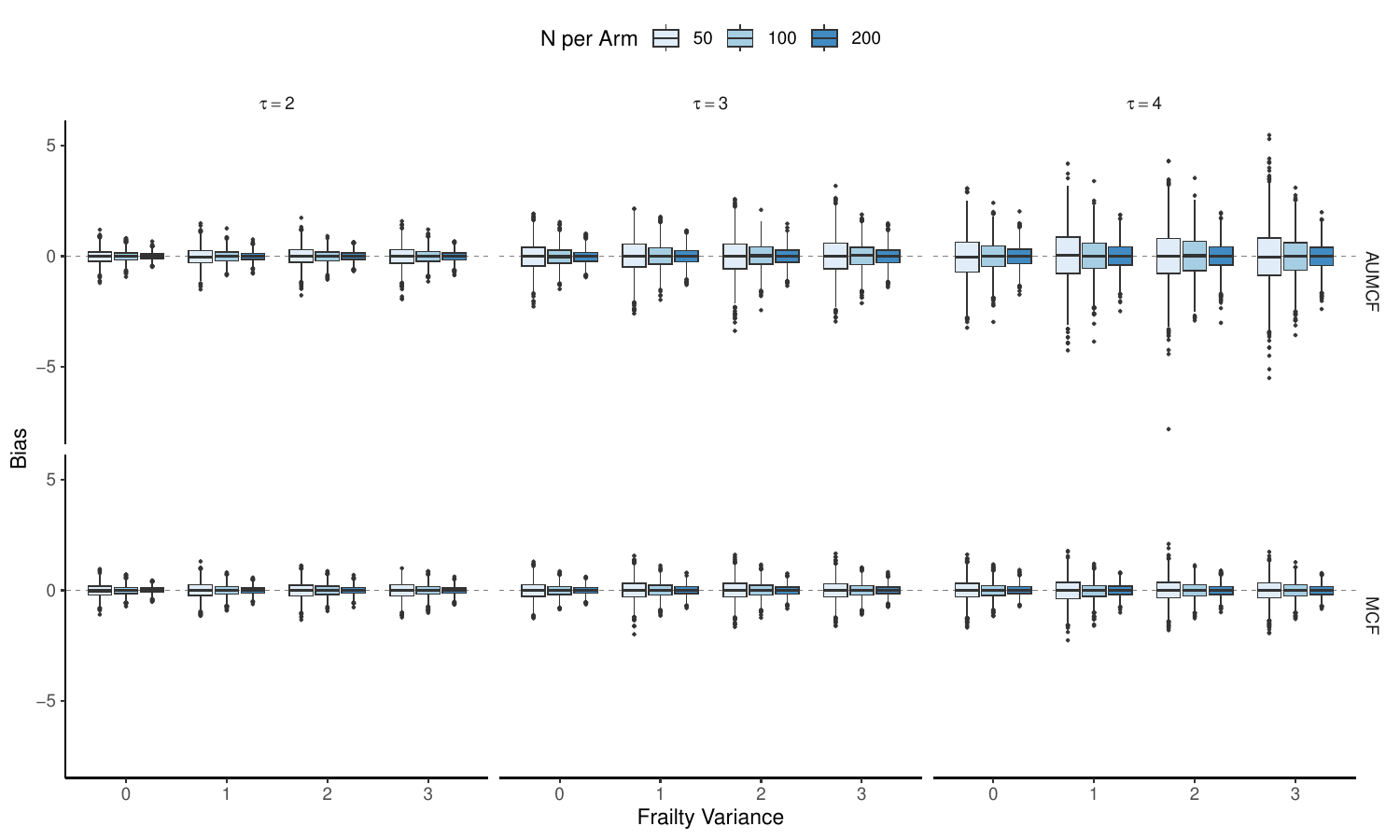}
    \end{subfigure}
    \begin{subfigure}{0.48\textwidth}
        \centering
        \caption{Coverage}
        \includegraphics[width=\textwidth]{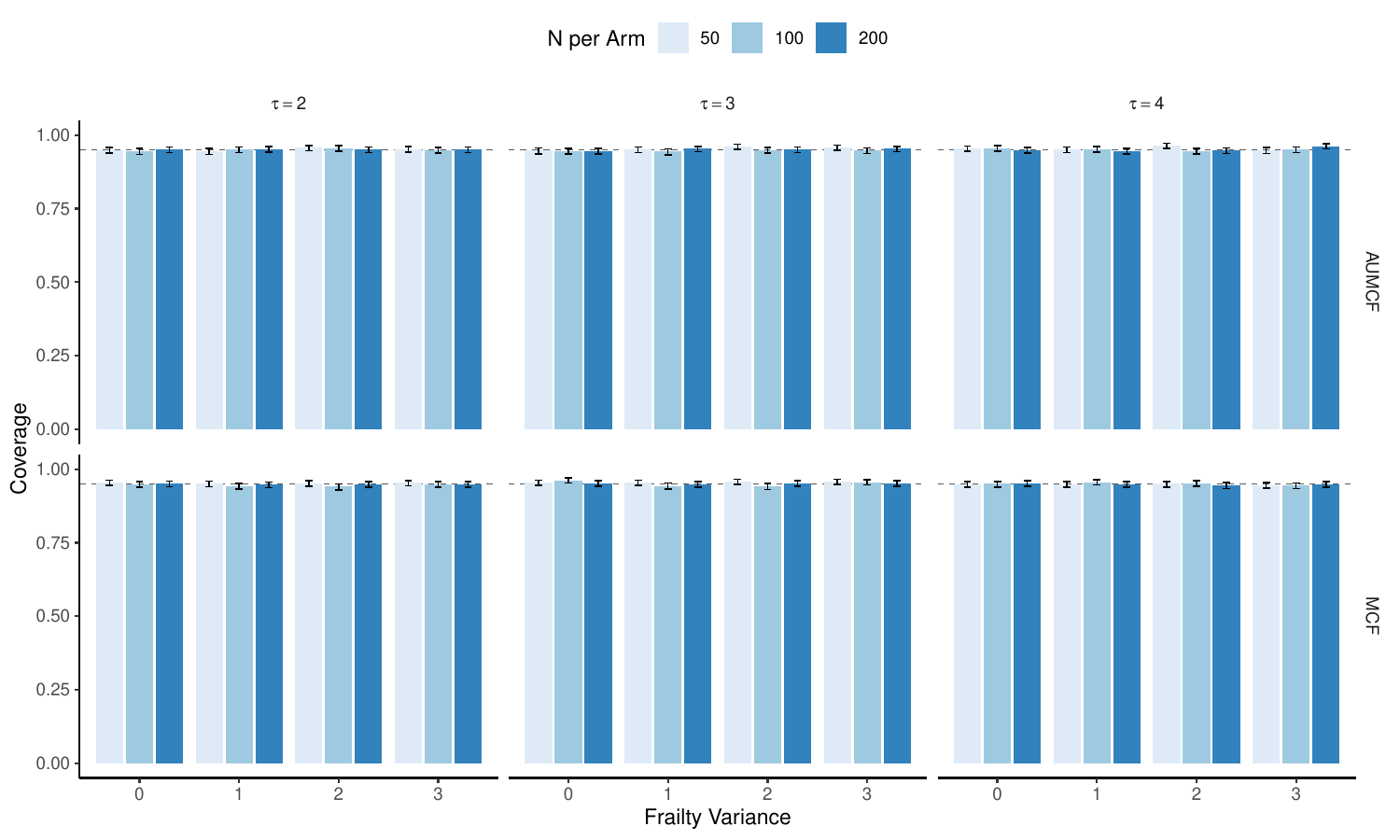}
    \end{subfigure}
    \caption{\textbf{Estimation accuracy and coverage are robust to dependence between death and recurrent events.} Bias and empirical 95\% confidence interval coverage are shown for the treatment effects on the MCF and AUMCF across sample sizes, truncation times, and shared-frailty variances. Larger frailty variances induce stronger dependence between the death and recurrent-event rates.}
    \label{fig:frailty}
\end{figure}

To investigate the impact of dependence between the death and recurrent-event rates, data were generated in the presence of a shared multiplicative frailty. The frailty variance was varied across the set $\sigma_{F}^{2} \in \{0, 1, 2, 3\}$, with increasing variance inducing stronger dependence. \textbf{Figure \ref{fig:frailty}} demonstrates that estimation accuracy and confidence interval coverage were insensitive to the presence of tight coupling between the death and recurrent-event rates. For concision, results are shown for the effects of treatment on the MCF ($\beta_{A}$) and AUMCF ($\gamma_{A}$) only; results for the other regression parameters were similar. 

\subsubsection{Covariate-dependent terminal event process}

We next evaluated performance when covariates affected the terminal-event process rather than the recurrent-event process. In these simulations, the recurrent-event rate was held fixed, while the treatment assignment, baseline covariate, or their interaction affected the death rate. Accuracy and coverage for the MCF and AUMCF are shown in \textbf{Figures \ref{supp:fig:terminal-event-beta}} and \textbf{\ref{supp:fig:terminal-event-gamma}}. Estimation error remained centered near zero for both the MCF and AUMCF regression targets across sample sizes, truncation times, and terminal-event effect sizes. Empirical coverage of 95\% confidence intervals was approximately nominal in all scenarios. These results support the validity of pseudo-value regression when covariate effects on cumulative recurrent-event burden operate through truncation by death.


\subsubsection{Statistical inference}

\begin{figure}
    \centering
    \includegraphics[width=0.8\textwidth]{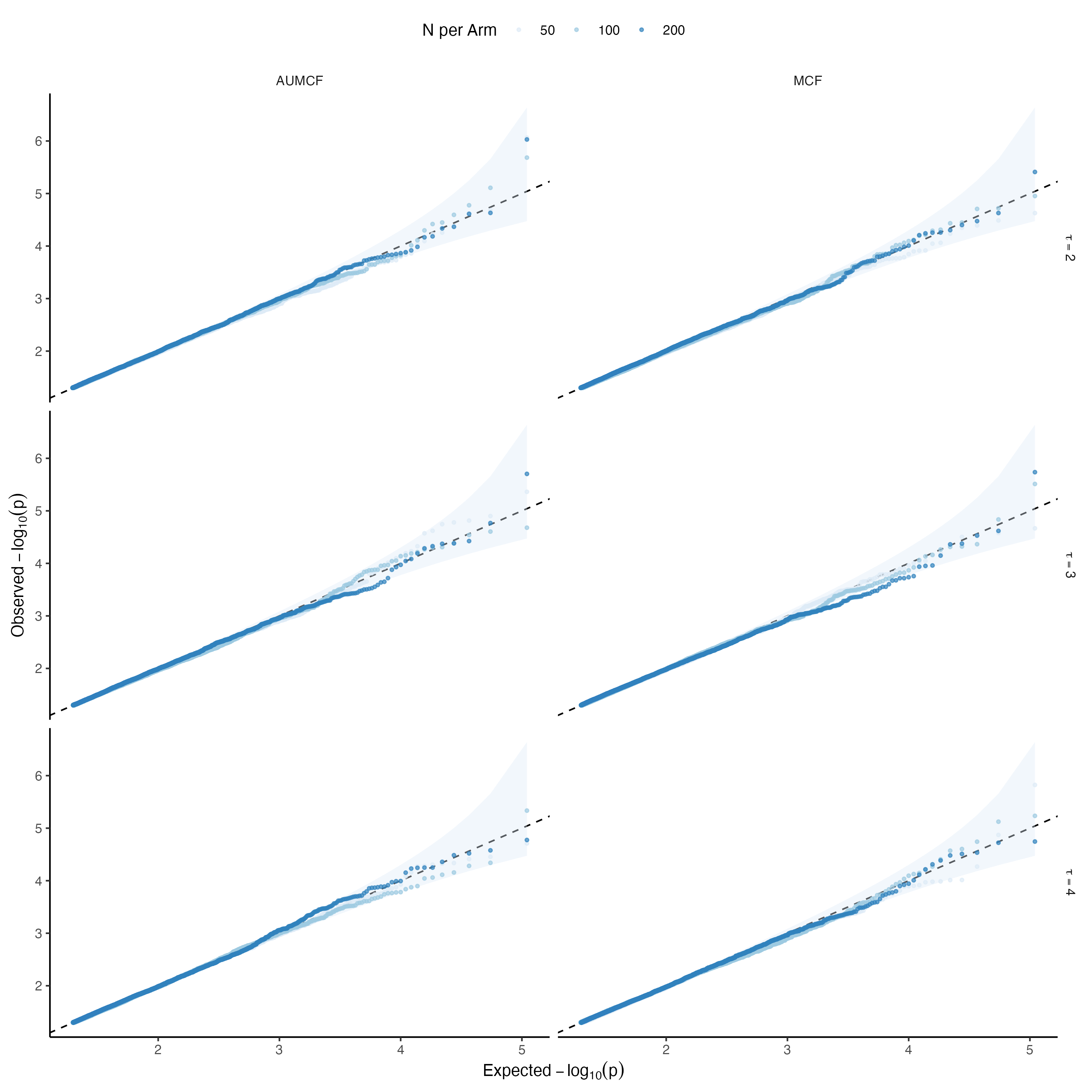}
    \caption{\textbf{Type I error is calibrated under the null.} Uniform quantile-quantile plots are shown for Wald-test p-values under the null hypotheses of no treatment effect on the MCF and AUMCF regression targets. Simulations were conducted in the absence of treatment effects on the recurrent-event rate, either directly or through interaction with the baseline covariate.}
    \label{fig:t1e}
\end{figure}

For type I error simulations, data were generated in the absence of any effect of treatment on the recurrent event rate, either directly or via the interaction. The target parameters were estimated via linear regression. Inference on the null hypotheses $H_{0}:\beta_{A} = 0$ and $H_{0}:\gamma_{A} = 0$ was performed via the standard Wald test. The uniform quantile-quantile plots in \textbf{Figure \ref{fig:t1e}} indicate that the resulting p-values are uniformly distributed under the null. Results for $10^{5}$ simulations are shown. For power simulations, the sample size per arm and the effect of treatment on the recurrent event rate were varied. Power is defined as the fraction of simulations in which the null hypothesis was correctly rejected. As expected, power increased with the sample size and the magnitude of the treatment effect. The MCF- and AUMCF-based tests had similar power for detecting the presence of a treatment effect. 

\begin{figure}
    \centering
    \includegraphics[width=\textwidth]{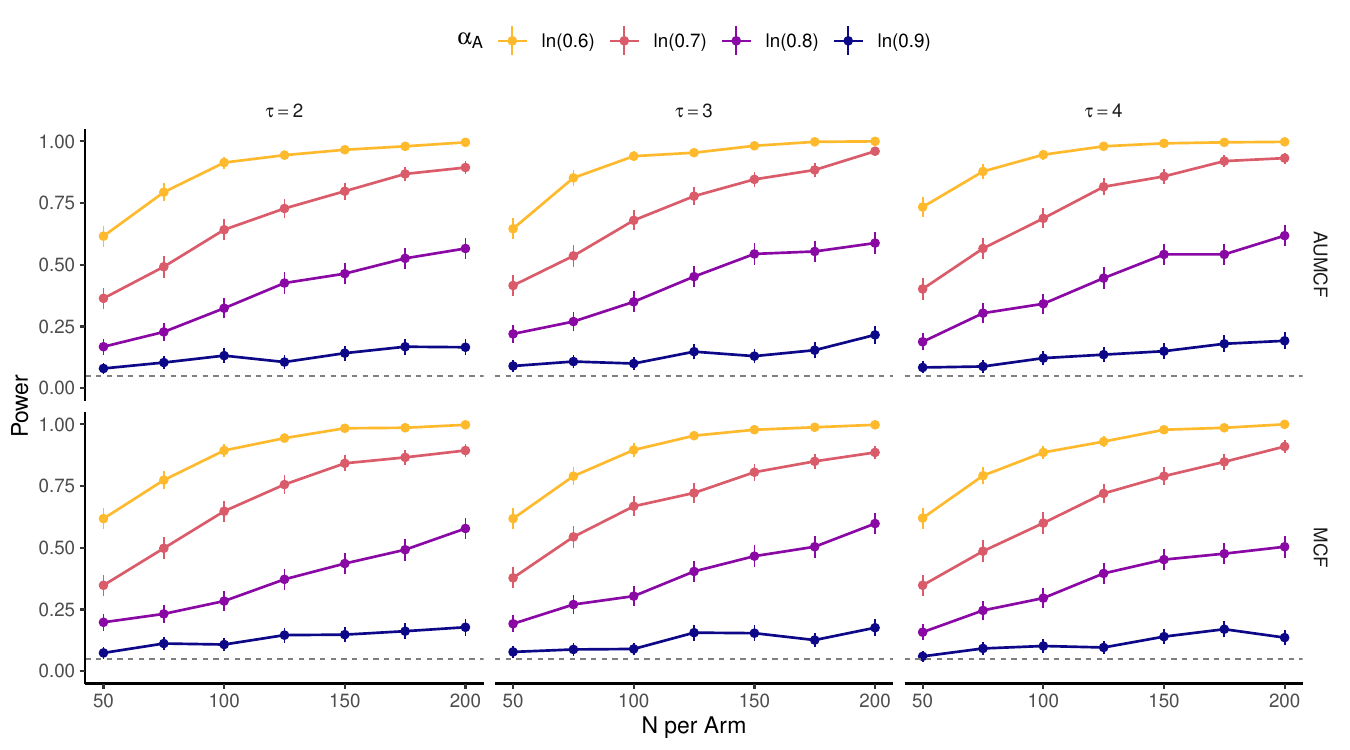}
    \caption{\textbf{Power increases with sample size and treatment effect magnitude.} Empirical power is shown for Wald tests of the treatment effects on the MCF and AUMCF regression targets across sample sizes, truncation times, and recurrent-event treatment-effect parameters. Power was defined as the proportion of simulations in which the null hypothesis of no treatment effect was rejected.}
    \label{fig:power}
\end{figure}

\clearpage 
\section{Data Application}
The ORATORIO study was a randomized phase III double-blind, placebo-controlled clinical trial in Primary Progressive Multiple Sclerosis (PPMS) \cite{montalban2016efficacy}. The trial implemented a 2:1 randomization ratio resulting in 488 patients randomized to ocrelizumab and 244 randomized to placebo. The goal of the study was to evaluate the efficacy and safety of ocrelizumab with 12-week confirmed disability progression as a primary endpoint. In this retrospective analysis, we demonstrate analyzing the composite Confirmed Disability Progression (cCDP), a multi-component time-to-event endpoint that considers three types of disability progression events: 1) confirmed disability worsening based on the Expanded Disability Status Scale (EDSS) score, 2) 20\% increase in the timed 25-foot walk test (T25FWT), or 3) 20\% increase in 9-hole peg test (9HPT). This multi-component outcome is typically analyzed in a time-to-event analysis where the event time is defined by the occurrence of the first of any of these three event types. In contrast, the MCF approach considered in this paper does not ignore events after the first; rather, it aggregates events across time, allowing the curve to jump multiple times for patients who experience multiple events. Thus, the MCF summarizes long-term disability accumulation in patients with MS. \\

We applied pseudo-value regression using baseline clinical and biomarker assessments known to be prognostic in MS. They are all baseline values of various clinical and biomarker assessments known to be important for multiple sclerosis patients that are standard collections in clinical trials. The set includes baseline EDSS, 9HPT, T25FWT, T2 lesion volume, baseline brain volume, and age. Importantly, three of these covariates are the baseline values of the clinical assessments used to define the eventual outcome.\\

\begin{figure}[H]
    \centering
    \includegraphics[width=0.75\linewidth]{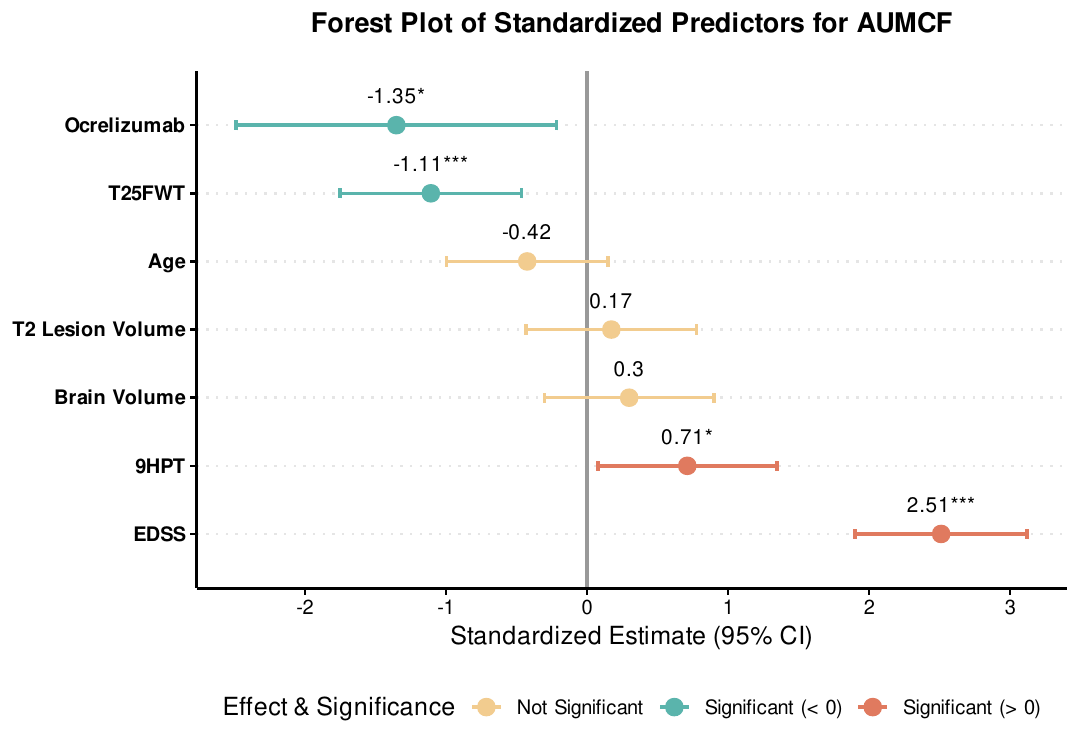}
    \caption{\textbf{Forest plot of pseudo-value based regression estimate for the covariate effects.} The covariates are standardized before putting them into the model, that is, the effects in the plot correspond to one standard deviation increase in the unit of the covariate. The label $^{***}$ corresponds to a p-value < 0.001,$^{**}$ < 0.01, and $^{*}$ < 0.05.}
    \label{fig:application}
\end{figure}

The covariate effects are visualized in a forest plot in Figure \ref{fig:application}. We can interpret these as the expected change in cumulative event-time burden due to a one unit (and in the model presented one standard deviation) increase in the covariate. We see that baseline EDSS score is a highly prognostic covariate. Baseline 9HPT time is also positively associated with cumulative disability burden. \\

We applied pseudo-value regression to the AUMCF to estimate covariate effects on cumulative disability burden. For instance, we can say that for a one standard-deviation increase of EDSS score, one expects a 2.51 year increase in disability-time accumulation. Incorporation of covariates also has the potential to increase efficiency. We note a 13.6\% reduction in the standard error (0.578 adjusted vs. 0.669 unadjusted)  of the treatment effect estimate due to covariate adjustment. The pseudo-value based standard errors for the treatment effect were smaller than those based on covariate adjustment via augmentation using the same set of covariates (0.578 vs. 0.632, respectively; 8.4\% reduction). As a sensitivity analysis, we considered a number of standard error estimators, which all yielded consistent inference for the reduction of disability event time on treatment. They are included in the supplementary materials.

\section{Discussion}

In this paper, we have developed and validated a pseudo-value regression framework for the MCF and AUMCF at a fixed truncation time. The approach allows estimation of covariate effects on cumulative recurrent-event burden without requiring a parametric model for the underlying recurrent-event intensity. Using influence-function-based pseudo-values avoids repeated leave-one-out estimation. Across the simulation settings considered, the resulting estimators were unbiased and confidence intervals achieved close to nominal coverage, including when recurrent and terminal events were dependent through shared frailty and when covariate effects operated through the terminal-event process. The power simulations further illustrate how the framework may be used to inform sample-size planning. \\

A practical advantage of the proposed approach is that, once the pseudo-values have been generated, regression can be performed using standard GEE machinery. Under an identity link, this reduces to fitting a linear model, for example with \texttt{lm()} in \texttt{R}. This is useful because the effects of covariates on the (AU)MCF generally do not have an obvious closed-form expression. When the working linear model is correctly specified, the regression coefficients have the usual conditional-mean interpretation. Otherwise, they represent the best linear projection of the conditional MCF or AUMCF onto the selected covariates. \\

From a clinical perspective, the proposed approach provides interpretable measures of covariate associations on the expected event-count or cumulative event-time scale. This allows associations to be assessed not only in terms of statistical significance, but also using clinically meaningful summaries such as the change in the expected number of events by a fixed time or the change in cumulative disease burden over follow-up. In observational settings, inclusion of covariates allows adjustment for measured confounders, enabling (AU)MCF regression to be incorporated into causal inference analyses under the usual identification assumptions. In randomized trials, covariate adjustment may improve the precision of estimated treatment effects as demonstrated in the ORATORIO study. Furthermore, the framework additionally supports more streamlined subgroup analyses through inclusion of interaction terms between treatment and subgroup indicators in linear pseudo-value models. This enables more sophisticated approaches, including Bayesian subgroup shrinkage analyses that can also be more easily implemented using pseudo-value regression models \cite{wolbers2026unified}. \\

A limitation of the proposed approach is that the (AU)MCF estimators are justified under non-informative censoring, which in the present implementation requires that censoring is independent of the recurrent-event process, terminal-event time, and baseline covariates included in the regression model. This assumption may be restrictive when dropout depends on observed prognostic characteristics. To accommodate covariate-dependent censoring, the framework could be extended by estimating the marginal functional of interest, either the MCF or AUMCF, using inverse-probability-of-censoring weights. Development and evaluation of this extension is a useful direction for future work. \\

Pseudo-value regression provides a simple and interpretable framework for analyzing recurrent-event and multi-component endpoints on the expected event-count or cumulative event-time scale. By enabling direct regression modeling of cumulative disease burden, the proposed approach supports a broader range of scientific and clinical questions than marginal treatment comparisons alone. 

\printbibliography

\end{document}